\documentclass{elsart}
\usepackage{epsfig}
\usepackage{amsmath}
\usepackage{amssymb}

\begin{document}

\begin{frontmatter}

\title{Similarity transformations for Nonlinear Schr\"odinger
Equations with time varying coefficients: Exact results.}

\author{V\'{\i}ctor M. P\'erez-Garc\'{\i}a}
\address{Departamento de Matem\'aticas, E. T. S. de Ingenieros
Industriales, \\
Universidad de Castilla-La Mancha 13071 Ciudad Real, Spain}

\author{Pedro J. Torres}
\address{Departamento de Matem\'atica Aplicada, \\
Universidad de
Granada, 18071 Granada, Spain.}

\author{Vladimir V. Konotop}
\address{Centro de F\'{\i}sica Te\'{o}rica e Computacional,
Universidade de Lisboa,
Complexo Interdisciplinar, Av. Prof. Gama Pinto 2, Lisbon 1649-003,
Portugal and Departamento de F\'{\i}sica,
Universidade de Lisboa, Campo Grande, Ed. C8, Piso 6, Lisboa
1749-016, Portugal}

\begin{abstract}
In this paper we use a similarity transformation connecting some
families of Nonlinear Schr\"odinger equations with time-varying
coefficients with the autonomous cubic nonlinear Schr\"odinger
equation. This transformation allows one to apply all known results
for that equation to the original non-autonomous case with the additional
dynamics introduced by the transformation itself. In particular,
using stationary solutions of the autonomous nonlinear
Schr\"odinger equation we can construct exact breathing  solutions
to multidimensional non-autonomous nonlinear Schr\"odinger equations. 
An application is given in which
we explicitly construct time dependent coefficients leading to
solutions displaying weak collapse in  three-dimensional
scenarios. Our results can find physical applicability in mean field models of Bose-Einstein condensates and in the field of
dispersion-managed optical systems.
\end{abstract}

\end{frontmatter}

\section{Introduction}

Nonlinear Schr\"{o}dinger (NLS) equations appear in a great array of
contexts \cite{Vazquez} as for example in optics of nonlinear media \cite{Kivshar}, photonics \cite{Hasegawa}, plasmas \cite{Dodd}, mean-field theory of Bose-Einstein condensates \cite{Book}, condensed matter physics, \cite{scott} and many other fields. In some of these fields and many others, the NLS equation appears as an asymptotic limit for a slowly varying dispersive wave envelope propagating in a nonlinear medium.

In the last years there has been a strong interest in different types
of NLS equations with cubic nonlinearity whose coefficients depend on the evolution variable. As the first example we mention \emph{stabilized solitons}, which appear
in multidimensional NLS equations when the nonlinear coefficient is controlled appropriately. These objects were predicted to exist in optical applications \cite{Berge}, but it was soon realized that a similar phenomenon arises in the context of mean field models of Bose-Einstein condensation (BEC) \cite{AB,ST,Gaspar}. 
As the second example, relevant to the  present paper, is {\em enhancement of the collapse} by varying nonlinearity~\cite{KP}. More references on this field of
activity can be found in Ref. \cite{Ft3}.
 
Another field of application where the coefficients of the NLS model
depend on the evolution variable is that of dispersion-managed
optical solitons. In that case it is the second derivative term the
one which is modulated along the longitudinal direction in an effort
to compensate the loss and gain dynamics (which is also incorporated
into a spatially-dependent nonlinear coefficient) (see e.g. the
review \cite{DM2}). The spatial aspects of the dynamics of these dispersion managed
solitons are another topic of recent interest \cite{Malo2}  where modulated coefficients arise.

Stimulated by Ref.~\cite{HW}, numerous studies of the one-dimensional NLS equation with combined temporal modulation of various coefficients, including also linear dissipation, have recently been reported~\cite{1D,Liang}, where the main attention was focused on properties of solitary wave solutions.

The aim of the present paper is to describe several two- and three-dimensional models which can be reduced to a standard NLS equation with constant coefficients by means of similarity transformations and describe some of their solutions concentrating on localized pulsating pulses, blowing up solutions as well as on the solutions which collapse in the case of time independent coefficients but are global in the case of varying coefficients.

\section{Similarity transformations for nonautonomous NLS equations}

\subsection{Model statement and equations for the transformation parameters}

To fix ideas we will denote the evolution variable as $t$, and
concentrate on the dynamics of solutions of the NLS equation
\begin{eqnarray}\label{NLS}
i \frac{\partial \psi}{\partial t} = -\frac{\alpha(t)}{2} \triangle
\psi + \frac{1}{2} \Omega(t) r^2 \psi + g(t)|\psi|^2 \psi - i\gamma(t) \psi.
\end{eqnarray} 
For the specific situation of applications of this model in the mean-field theory of BEC we have added a potential depending quadratically on the spatial variables ($r = |x|$). We will consider complex solutions of Eq. (\ref{NLS})
defined on $\mathbb{R}^d$, i.e. $x\in \mathbb{R}^d$,  satisfying initial data  $\psi(x,0)=\psi_0(x)$. 

Let us now make the similarity transformation
\begin{eqnarray}\label{simm}
\psi(x,t) = \frac{1}{\ell(t)} e^{if(t) r^2}
u\left(\frac{x}{L(t)},\tau(t)\right),
\end{eqnarray}
where $\ell(t), L(t), f(t)$ and $\tau(t)$ are scaling functions to be fixed later. 

This transformation is analogous to the so-called \emph{lens
transformation} \cite{Sulem} which has been also used in other BEC
problems \cite{Rybin,KevFran} and differs (mainly because of the
extra freedom provided by the definition of the new time $\tau$) from
the transformations which are frequent in the context of
dispersion-managed solitons \cite{HW}. For simplicity and without
loss of generality we choose $t_0 = 0$, $\ell(0) = 1$, $L(0)=1$ and
$\tau(0) = 0$. 

In this paper we will use the nonlinear rescaling group given by Eq. (\ref{simm}) to transform the time dependent problem given by Eq. (\ref{NLS}) into simpler, time independent ones by choosing appropriately the transformation parameters. This transformation has 
 been used in the study of linear Schr\"odinger equations in Ref. \cite{Soler} but in the opposite way, i. e. to tranform the linear Schr\"odinger equation into a time-dependent problem whose asymptotic behavior provides information on the autonomous one via rescaling.

To determine the scaling functions we require the function $u=u(\eta,\tau)$, where $\eta=x/L(t)$ to solve the NLS equation with the stationary coefficients
\begin{eqnarray}\label{simple}
i \frac{\partial u}{\partial \tau} = - \frac{1}{2} \triangle_{\eta} u,
+ \sigma |u|^2u
\end{eqnarray}
 where $\sigma = \pm 1$ (the choice of $\sigma$ may imply a redefinition of the sign of $g$) 
 and the initial data for Eq. (\ref{simple})
are given by $u(\eta,0) = e^{-if_0 r^2}\psi_0(\eta)$, being
$\eta = x$ and $f_0=f(0)$.
Then, inserting Eq. (\ref{simm}) into Eq. (\ref{NLS}) leads to the
following equations for the scaling parameters
\begin{subequations}
\label{sima}
\begin{eqnarray}
\frac{d \ell}{dt} & = & \alpha(t)  fd \ell+\gamma(t)\ell, \label{simaa} \\
\frac{df}{dt} & = & - 2\alpha(t) f^2 - \frac{1}{2} \Omega(t),
\label{simab}\\
\frac{dL}{dt} & = & 2 \alpha(t) f L, \label{simac}\\
\frac{d\tau}{dt} & = & \frac{\alpha(t)}{L^2},  \label{simad}\\
\frac{\alpha(t)}{L^2} & = & \sigma \frac{g(t)}{\ell^2}. \label{simae}
\end{eqnarray}
\end{subequations}

What we have now in Eq. (\ref{simple}) is the autonomous cubic
NLS equation. Thus, the transformation given
by Eqs. \eqref{sima} allows one to study the properties of solutions
of Eq. \eqref{NLS} from those of the simpler, cubic, nonlinear
Schr\"odinger equation given by Eq. (\ref{simple}) for which many
results are available (see e.g. \cite{Sulem}), provided the coefficients are such that
they satisfy Eqs. (\ref{sima}). In particular we can construct
explicit solutions of Eq. (\ref{NLS}) from those of Eq.
(\ref{simple}).

An obvious application is to start with stationary solutions of
Eq. (\ref{simple}) which are known to exist for $\sigma = -1$
\cite{Sulem} and from them construct breathing and blowing up solutions of Eq.
(\ref{NLS}). In this case, all the time dynamics of the solutions,
which are self-similar, is contained into Eqs. (\ref{sima}). We will present an explicit example of this methodology later.

Another application, valid specially for the case of two spatial dimensions, is to use the method of moments
in the framework of Eq. (\ref{simple}) to compute the evolution of some relevant integral quantities in the non-autonomous case.
Finally, explicit soliton and multisoliton solutions
of Eq. (\ref{simple}) for $d=1$ can be used to construct solutions of Eq. (\ref{NLS}) with more complex dynamics in the case of coefficients
varying in time, a fact which has been partially explored in the field of dispersion-managed solitons.

These are only a few illustrations of what can be expected from
the application of our transformations to specific situations
using them as a link between Eqs. (\ref{NLS}) and Eq.
(\ref{sima}).

In what follows we concentrate on describing the possible dynamics
which can be obtained from Eqs. (\ref{sima}) with the
understanding that the full dynamics in the framework of Eq.
(\ref{NLS}) will be a combination of the dynamics of $u$ given by
Eq. \eqref{simple} with the dynamics of the scaling parameters
given by Eq. (\ref{sima}).

\subsection{Solution of the equations for the scaling parameters}

The solution of Eqs. (\ref{simaa}) and (\ref{simac}) is immediate
\begin{subequations}
\label{solas}
\begin{eqnarray}
\ell(t)  & = & \Gamma(t)\exp\left(d \int_0^t \alpha(t') f(t') dt'\right), \\
L(t)  & = & \exp\left(2 \int_0^t \alpha(t') f(t') dt'\right),
\end{eqnarray}
\end{subequations}
where
\begin{eqnarray}
	\Gamma(t)=\exp\left(\int_0^t\gamma(t')dt'\right),
\end{eqnarray}
and thus obviously $\ell(t) = \Gamma(t)L(t)^{d/2}$ which, together with Eq.  (\ref{simae}) implies that 
\begin{eqnarray}
\label{rest}
g(t) =\sigma  \alpha(t)\Gamma(t)^2 L(t)^{d-2}.
\end{eqnarray}
This means that the transformations proposed allow us to obtain
solutions for any of the functions $\alpha(t)$, $g(t)$, and $\gamma(t)$, but once
two of them are chosen the third one is fixed by \eqref{rest}.
Although this change is only valid for a limited number of
situations we wish to remind that in those remarkable cases the
information we get relating solutions of the explicitly
time-dependent problem with to those of the autonomous Eq.
(\ref{simple})  is very complete.

To proceed further we need to solve Eq. (\ref{simab}) since, using
Eqs. (\ref{solas}), it would give us the values of $L(t), \ell(t)$.
Unfortunately the general solution of Eq. (\ref{simab}) cannot be
written explicitly except for very special cases.

We will  concentrate first on several specific examples which are specially interesting for applications and will return
later to the full variety of behaviors which are possible within the framework of Eqs. (\ref{sima}).

\section{Applications}

\subsection{Systems without external potentials $\Omega(t) \equiv 0$.}

This situation is typical of problems arising in nonlinear optics
(i.e. dispersion-managed systems) in which $\alpha(t)$ is a periodic
piecewise constant function of the propagation variable (which is
usually denoted $z$ because of its physical interpretation as a propagation distance).

Let us first consider a dissipationless case when $\Omega(t) = 0$, Eq. (\ref{simab}) can be integrated to get
\begin{eqnarray}
\label{f}
f(t) = \frac{f_0}{2f_0 \int_0^t \alpha(t') dt' +1}.
\end{eqnarray}
Substituting this expression into Eqs. (\ref{solas}) we get the
solution in the form of quadratures.

A typical problem in the context of dispersion-managed systems is the
existence of periodic solutions. Limiting the consideration to the case where  
$\left| 2f_0 \int_0^t \alpha(t') dt' \right| <1$ and using explicit forms of the solutions we get that in order to satisfy the periodicity condition
$L(T) = L(0)$ the following condition must hold
\begin{eqnarray}
\int_0^T \frac{\alpha(t)}{1+2f_0\int_0^t \alpha(t') dt'} dt  = 0.
\end{eqnarray}
Using the change of variables $\beta(t) = f_0 \int_0^t \alpha(t')
dt'$ this condition can be seen to be equivalent to $\log \left(1+
2\beta(T)\right) = 0$ and then $\beta(T) = 0$ which leads to
\begin{eqnarray}
\label{average}
\int_0^T \alpha(t') dt' = 0.\label{zeroav}
\end{eqnarray}
In particular, this result implies that under the assumption $\Omega(t) = 0$ any  function $\alpha(t)$
periodic and with zero average will lead  to periodic $L(t), \ell(t)$ and
$f(t)$. From now on, we assume that $T$ is the smallest period of
$\alpha(t)$.

In a general case using (\ref{f}) we can integrate Eq. (\ref{simad}) to obtain
\begin{eqnarray}
\label{tau}
\tau(t) = \frac{\int_0^t \alpha(t') dt'}{2f_0\int_0^t \alpha(t') dt'+1}.
\end{eqnarray}
Subject to the constrain (\ref{average}) we have that $\tau(T) = 0$ and $\tau(t)$ is bounded, i.e  $\tau_m \leq \tau(t) \leq \tau_M$  with $\tau_m\tau_M<0$. 

\subsection{Systems without dispersion management
$\alpha(t)\equiv 1$}\label{subsect3.3}

This situation arises physically in Feschbach-resonance managed Bose-Einstein
condensates for which $g(t)$ can be continuously controlled
but the dispersion is fixed $\alpha(t)=1$. Obviously, in the case $d=2$ 
 $g = \sigma \Gamma(t)^2 $ and sources of nontrivial dynamics are the trapping potential and dissipative term depending on time.
 
In all situations with $\alpha = 1$, the relevant equation to
solve is the Riccati equation (\ref{simab}) which in our case reads
\begin{eqnarray}\label{ric2}
\frac{df}{dt}  =  - 2 f^2 - \frac{1}{2} \Omega(t).
\end{eqnarray}
For arbitrary dimension $d$, some general information can be
obtained from classical tools from the qualitative theory of
ordinary differential equations. For instance, if $\Omega$ is
$T$-periodic and negative, then there exists two (real) periodic
solutions of Eq. (\ref{simab}), one of them positive and the other
negative. This is easily deduced from the classical theory of
upper and lower solutions \cite{O}. This sign information can be
used to get the qualitative asymptotic behaviour of $\ell,L$ from
Eq. (\ref{solas}). For instance, if $f$ is chosen positive then
\begin{subequations}
\begin{eqnarray}
 \lim_{t\to+\infty}\ell(t)=\lim_{t\to+\infty}L(t)=+\infty,
 \end{eqnarray}
whereas if $f$ is negative then
\begin{eqnarray}\label{L2}
\lim_{t\to+\infty}\ell(t)=\lim_{t\to+\infty}L(t)=0.
\end{eqnarray}
\end{subequations}
Additional information can be provided by writing $\Omega(t)=\lambda+\tilde\Omega(t)$,
where $\int_0^T\tilde\Omega(t)dt=0$ and $\lambda$ is considered as a
parameter. Then, a typical Ambrosetti-Prodi result holds, that is,
there exists a Hopf bifurcation for a critical value $\lambda_0<0$, such
 for $\lambda<\lambda_0$ there exist two periodic solutions, while for  $\lambda>\lambda_0$
 no periodic solutions exist. Finally, for the critical value $\lambda=\lambda_0$, there is exactly a periodic
solution.

When $\Omega$ is a general function (not
necessarily $T$-periodic), the following
reasoning is at hand. If $x$ is a real solution of the linear
equation
\begin{eqnarray}\label{linear}
\ddot x+\Omega(t)x=0,
\end{eqnarray}
 then it is a simple substitution to check that $f(t)=\dot x(t)/2x(t)$ is a (real) solution of
Eq. (\ref{ric2}). Note that from (\ref{solas}) we get
$\ell(t)=\left[x(t)/x(0)\right]^{d/2}$, $L(t)=x(t)/x(0)$.  Eq. (\ref{linear}) is the well-known
Hill equation, which has been widely studied \cite{MW}. If $\Omega$
is $T$-periodic, then Eq. (\ref{linear}) is oscillatory (i.e., every
solution has an infinite number of zeroes going to $\pm\infty$) if
and only if the mean value of $\Omega$ is positive. In this case, the
solution $f(t)=\dot x(t)/2x(t)$ is only defined in a finite
interval $(t_0,t_1)$ where $t_0,t_1$ are two consecutive zeroes of
$x$. As a consequence,
 \begin{eqnarray}\label{L1}
 \lim_{t\to t_1}\ell(t)=\lim_{t\to t_1}L(t)=0.
 \end{eqnarray}

As the first practical output of the above analysis we consider an ``explicit" example of a pulsating solution in the critical dimension $d=2$. Indeed, let us consider  
\begin{eqnarray}
\label{omega1}
	\Omega(t)=m(1-2\,\mbox{sn}^2(t,m)),
\end{eqnarray}
where sn$(x,m)$ is a Jacobi elliptic function.
A solution of Eq. (\ref{linear}) can be constructed explicitly in the form
\begin{eqnarray}
	x(t)=\mbox{dn}(x,m),
\end{eqnarray}
thus $x(t)$ is a positive periodic function. 
Let also $U$ be a solution of 
\begin{eqnarray}
	 \frac{1}{2} \triangle_{\eta} U -U
+ U^3=0,
\end{eqnarray}
with $U(\eta) \rightarrow 0$ when $|\eta| \rightarrow \infty$ 
and $u(\eta, \tau)=U(\eta)e^{i\tau}$. Such kind of solutions is known to exist \cite{Sulem}. For instance one of them is the so-called ``ground state" (or Townes soliton) which is a 
nodeless solution decaying exponentially at infinity. Then, the function
\begin{eqnarray}
	\psi=\frac{1}{\mbox{dn}(t,m)}U\left(\frac{x}{\mbox{dn}(t,m)}, \tau(t)\right)e^{i\tau(t)}\exp\left(ir^2\frac{m\mbox{cn}(t,m)\,\mbox{sn}(t,m)}{2\,\mbox{dn}(t,m)}\right),
\end{eqnarray}
where $\tau(t)=\int_0^{t}\mbox{dn} (t',m)^{-2}dt'$, is a periodically pulsating solution solution of the equation
\begin{eqnarray}
\label{NLS1}
i \frac{\partial \psi}{\partial t} = -\frac{1}{2} \triangle
\psi + \frac{m}{2}(1-2\,\mbox{sn}^2(t,m)) x^2 \psi - |\psi|^2 \psi.
\end{eqnarray}
 
Another practical implication, which can be obtained from the above analysis, is related to collapse in three-dimensional scenarios
 for any of the two situations described by either Eqs. \eqref{L2} or \eqref{L1} (with the difference, that in the first case collapse happen at infinite time while in the second case it occurs at finite time). To this end, first we notice that either in a dissipationless case or in the case of a periodically varying dissipaion with zero mean value:  i.e when $\int_0^T\gamma(t)dt \equiv 0$, if $d=3$ and because of Eq. (\ref{rest})
 $L(t) \rightarrow 0$ implies $g(t) \rightarrow 0$. Moreover
 we can take as initial data any stationary solution of Eq. (\ref{simple})
 (i.e. the ground state \cite{Sulem}) when $\sigma = -1$. Then we have self-similar collapsing solutions with decaying
 nonlinear coefficient  given by Eq. \eqref{rest}. In this situation what we have is an example of a three-dimensional scenario
 with a decaying nonlinearity for which the initial data collapses as a whole, i.e. an example of induction of weak collapse
  in a three-dimensional scenario. This leads to a situation characteristic of two-dimensional scenarios for which the dynamics is  ``less singular" than the usual strong collapse scenarios happening in three-dimensional problems with cubic nonlinearities. The smoother behavior of the situation is induced by the decay of the nonlinearity near the singularity. 
  
\section{Some general results  for sign definite $\alpha(t)$}

In this subsection we consider we general case under the restriction of definite $\alpha(t)\ne 0$ for any $t$.
Again, upper and lower solutions together with standard arguments from the theory of
ordinary differential equations provide some useful information.

If $\alpha,\Omega$ are periodic of the same period  and
$\alpha(t)\Omega(t)<0$ for all $t$ (that is, the respective signs
are constant and opposite) then there exists two (real) periodic
solutions of Eq. (\ref{simab}), one of them positive and the other
negative. Also a Hopf bifurcation like in the previous case holds,
where the sign of $\alpha$ determines the direction of such a
bifurcation.

Other line of procedure is the following. The change in the
dependent variable
\begin{eqnarray}\label{ch}
\varphi=-\frac{\dot\alpha}{4\alpha}+\alpha(t)f,
\end{eqnarray}
transforms the Riccati equation $(\ref{simab})$ into the simpler
one
\begin{eqnarray}\label{ric3}
\frac{d\varphi}{dt}  = -2\varphi^2 -\frac{1}{2} \hat\Omega(t),
\end{eqnarray}
with
\begin{eqnarray}
\hat\Omega(t)=\alpha(t)\Omega(t)+\frac{1}{4}\frac{d}{dt}\left(\frac{\dot\alpha}{\alpha}\right).
\end{eqnarray}
This transformation allows to reduce the analysis to the case of constant dispersion
$\alpha(t) = 1$ which was discussed in Sec. \ref{subsect3.3}

 Finally, we want to discuss how, given an arbitrary  $T$-periodic coefficient
$\alpha(t)$, we can design $\Omega(t),g(t)$ in order to get
periodic responses as those obtained in the system without
external potential. Looking to Eq. (\ref{simad}), one realizes
that a necessary condition for the existence of a $T$-periodic
$\tau(t)$ is that $\alpha(t)$ must change its sign. This case
cannot be treated with the previous arguments, but we can use an
inverse procedure. Take $L$ a $T$-periodic function such that
 \begin{eqnarray}
 \int_0^T \frac{\alpha(t)}{L^2(t)}dt=0.
\end{eqnarray}
 Then $\tau(t)$ is $T$-periodic. Once $L$ is fixed, $f$ is determined
by Eq. (\ref{simac}) and finally $\Omega$ is given by Eq. (\ref{simab}).
Of course, $g$ is always fixed by the restriction Eq. (\ref{rest}).

 Finally, if we are not interested in the periodicity of $\tau(t)$
but only in the rest of coefficients, a similar trick is at hand.
Given an arbitrary $\alpha$, take $f$ such that $\int_0^T
\alpha(t)f(t)dt=0$. This guarantees the existence of periodic solutions
$\ell,L$ of Eqs. (\ref{simaa})-(\ref{simab}) (in fact all the
solutions of these equations are periodic) and again $\Omega$ is
determined by  Eq. (\ref{simab}).

\subsection{Systems with quasiperiodic coefficients.}

Let us suppose that the coefficients $\alpha,\Omega$ are periodic
with different period or, more generally, quasiperiodic functions.
Assuming again that $\alpha(t)\ne 0$ for any $t$, the change given
by (\ref{ch}) leads again to the Riccati equation (\ref{ric3})
with a quasiperiodic coefficient $\hat\Omega(t)$. Then, the
general theory of almost-periodic equations \cite{Kras} can be
used in order to prove a Hopf bifurcation as in Subsection
\ref{subsect3.3}. The analogous of the mean value for a
quasiperiodic function $p(y)$ is the generalized mean value
\begin{eqnarray}
M[p]=\lim_{T\to\infty}\frac{1}{2T}\int_{-T}^{T}p(t)dt.
\end{eqnarray}
Again, we parametrize $\hat\Omega(t)=\lambda+p(t)$ with $M[p]=0$
and then there exists $\lambda_0<0$, such that there exist two
quasiperiodic solutions for $\lambda<\lambda_0$ and no
quasiperiodic solutions for $\lambda>\lambda_0$. In this situation
the information on the critical value $\lambda=\lambda_0$ is lost,
and the only result remaining is the existence of  a bounded
solution.

\section{Conclusions}

In this paper we have constructed similarity transformations connecting some
families of Nonlinear Schr\"odinger equations with time-varying coefficients
with the autonomous cubic nonlinear Schr\"odinger equation.

These similarity transformations hold when specific conditions
linking the modulation of the dispersion, potential and
nonlinearity are satisfied. In those cases they can provide
valuable information since
 they allow to apply all known results for the autonomous cubic nonlinear Schr\"odinger equation to the non-autonomous case.

In particular, using stationary solutions of the autonomous nonlinear Schr\"odinger equation we have discussed how to construct
exact breathing solutions to the non-autonomous nonlinear Schor\"odinger equation.

In addition to discussing the dynamics which can be expected in several cases we have provided an specific
 application  in which  we explicitly construct time dependent coefficients leading to solutions displaying weak
 collapse in  three-dimensional scenarios, which is a very unexpected result. i.e. we obtain a family of collapsing solutions with a
 behavior which is characteristic of two-dimensional
 scenarios in a three-dimensional setting by taking a time-dependent nonlinear coefficient which vanishes at the collapse point.

 Finally let us comment that an extension
of the model equation in which only $p<d$ of the spatial variables
are modulated, i.e. with spatial derivatives of the form $\alpha(t)\sum_{j=1}^p\partial^2/\partial x_j^2 + \sum_{j=p+1}^d\partial^2/\partial x_j^2$ instead of the Laplacian operator, can
 also be treated following the same methodology presented here. However, the analysis becomes more complicated and the richness of scenarios
 is drastically reduced in relation to what is found here since the unmodulated derivatives impose strict conditions on the possible variations for $g(t)$.

Our results can find physical applicability in specific examples of mean field models of
Bose-Einstein condensates and in the field of dispersion-managed optical systems.

\textbf{Acknowledgements} This work has been supported by grants:
BFM2003-02832 and MTM2005-03483 (Ministerio de Educaci\'on y
Ciencia, Spain), PAI-05-001 (Consejer\'{\i}a de Educaci\'on y
Ciencia de la Junta de Comunidades de Castilla-La Mancha, Spain)
and POCI/FIS/56237/2004 (the Funda\c{c}\~ao para a Ci\^encia e a Tecnologia, Portugal and European program FEDER).

\end{document}